\documentclass[aip,jcp,reprint,noshowkeys,superscriptaddress]{revtex4-1}
\usepackage{graphicx,dcolumn,bm,xcolor,microtype,multirow,amscd,amsmath,amssymb,amsfonts,physics,wrapfig,txfonts,siunitx}
\usepackage[version=4]{mhchem}

\newcommand{\ie}{\textit{i.e.}}

\usepackage[normalem]{ulem}

\newcommand{\SupInf}{\textcolor{blue}{Supporting Information}}

\newcommand{\QP}{\textsc{quantum package}}

\newcommand{\Ndet}{N_\text{det}}
\newcommand{\Nbas}{N}

\usepackage[
	colorlinks=true,
    citecolor=blue,
    breaklinks=true
	]{hyperref}
\begin{document}

\newcommand{\LCPQ}{Laboratoire de Chimie et Physique Quantiques (UMR 5626), Universit\'e de Toulouse, CNRS, UPS, France}

\title{Hierarchy Configuration Interaction: Combining Seniority Number and Excitation Degree}

\author{F\'abris Kossoski}
\email{fkossoski@irsamc.ups-tlse.fr}
\affiliation{\LCPQ}
\author{Yann Damour}
\affiliation{\LCPQ}
\author{Pierre-Fran\c{c}ois Loos}
\email{loos@irsamc.ups-tlse.fr}
\affiliation{\LCPQ}

\begin{abstract}
{\bf Abstract:} 
We propose a novel partitioning of the Hilbert space, hierarchy configuration interaction (hCI), 
where the excitation degree (with respect to a given reference determinant) and the seniority number (\ie, the number of unpaired electrons) are combined in a single hierarchy parameter.
The key appealing feature of hCI is that each hierarchy level accounts for all classes of determinants whose number share the same scaling with system size.
By surveying the dissociation of multiple molecular systems, we found that the overall performance of hCI usually exceeds or, at least, parallels that of excitation-based CI.
For higher orders of hCI and excitation-based CI, 
the additional computational burden related to orbital optimization usually do not compensate the marginal improvements compared with results obtained with Hartree-Fock orbitals.
The exception is orbital-optimized CI with single excitations, a minimally correlated model displaying the qualitatively correct description of single bond breaking,
at a very modest computational cost.
\bigskip
\begin{center}
        \boxed{\includegraphics[keepaspectratio,width=2in]{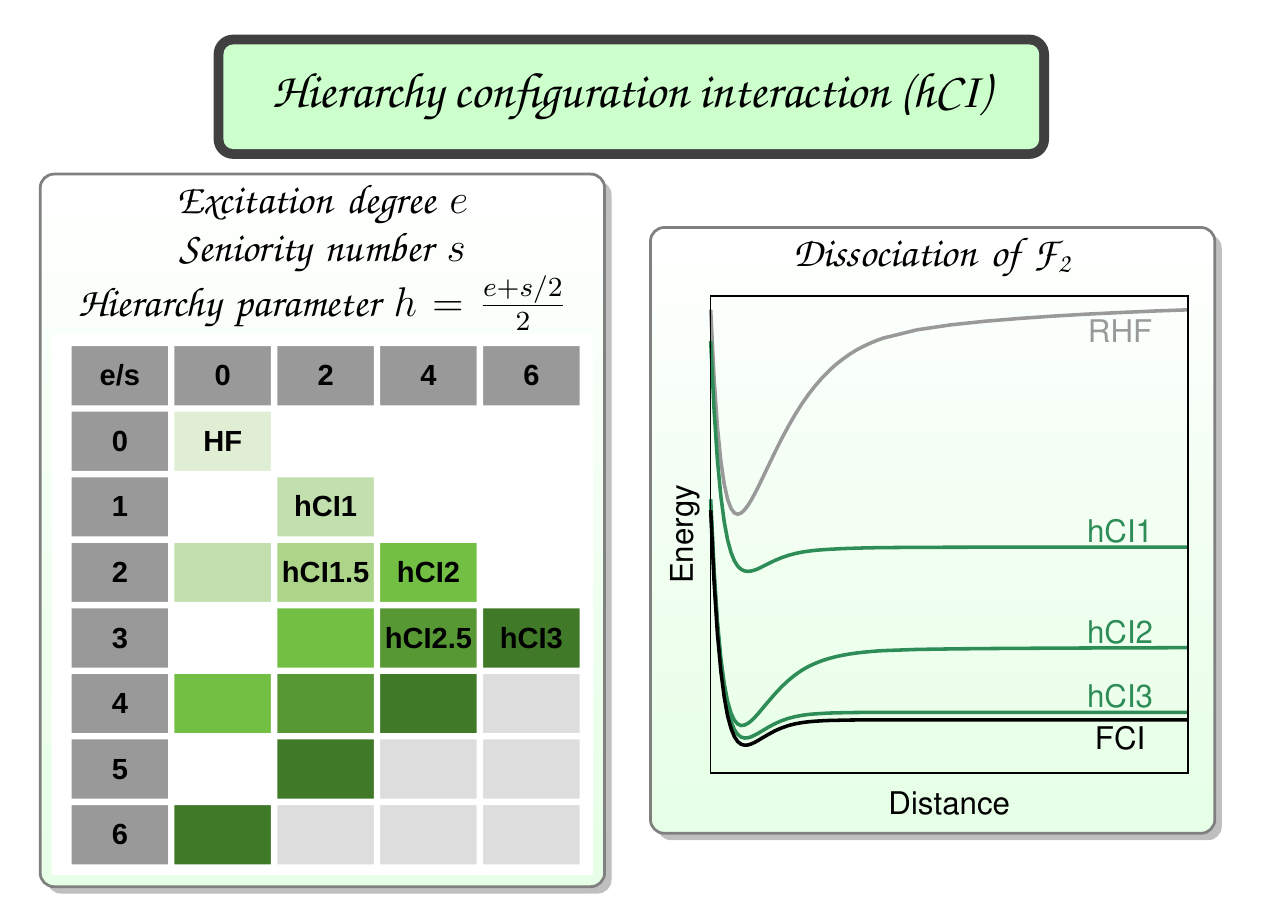}}
\end{center}
\bigskip
\end{abstract}

\maketitle


In electronic structure theory, configuration interaction (CI) methods allow for a systematic way to obtain approximate and exact solutions of the electronic Hamiltonian,
by expanding the wave function as a linear combination of Slater determinants (or configuration state functions). \cite{SzaboBook,Helgakerbook}
At the full CI (FCI) level, the complete Hilbert space is spanned in the wave function expansion, leading to the exact solution for a given one-electron basis set.
Except for very small systems, \cite{Knowles_1984,Knowles_1989} the FCI limit is unattainable, and in practice the expansion of the CI wave function must be truncated.
The question is then how to construct an effective and computationally tractable hierarchy of truncated CI methods
that quickly recover the correlation energy, understood as the energy difference between the FCI and the mean-field Hartree-Fock (HF) solutions.

Excitation-based CI is surely the most well-known and popular class of CI methods.
In this context, one accounts for all determinants generated by exciting up to $e$ electrons from a given reference, which is usually the HF determinant, but does not have to.
In this way, the excitation degree $e$ defines the following sequence of models:
CI with single excitations (CIS), CI with single and double excitations (CISD), CI with single, double, and triple excitations (CISDT), and so on.
Excitation-based CI manages to quickly recover weak (dynamic) correlation effects, but struggles in strong (static) correlation regimes.
It also famously lacks size-consistency which explains issues, for example, when dissociating chemical bonds.
Importantly, the number of determinants $\Ndet$ (which is the key parameter governing the computational cost, as discussed later) scales polynomially with the number of basis functions $\Nbas$ as $\Nbas^{2e}$.
 
Alternatively, seniority-based CI methods (sCI) have been proposed in both nuclear \cite{Ring_1980} and electronic \cite{Bytautas_2011} structure calculations.
In short, the seniority number $s$ is the number of unpaired electrons in a given determinant.
By truncating at the seniority zero ($s = 0$) sector (sCI0), one obtains the well-known doubly-occupied CI (DOCI) method, \cite{Bytautas_2011,Allen_1962,Smith_1965,Veillard_1967}
which has been shown to be particularly effective at catching static correlation,
while higher sectors tend to contribute progressively less. \cite{Bytautas_2011,Bytautas_2015,Alcoba_2014b,Alcoba_2014}
In addition, sCI0 is size-consistent, a property that is not shared by higher orders of seniority-based CI.
However, already at the sCI0 level, $\Ndet$ scales exponentially with $\Nbas$, since excitations of all degrees are included.
Therefore, despite the encouraging successes of seniority-based CI methods, their unfavorable computational scaling restricts applications to very small systems. \cite{Shepherd_2016}
Besides CI, other methods that exploit the concept of seniority number have been pursued. \cite{Limacher_2013,Limacher_2014,Tecmer_2014,Boguslawski_2014a,Boguslawski_2015,Boguslawski_2014b,Boguslawski_2014c,Johnson_2017,Fecteau_2020,Johnson_2020,Johnson_2022,Fecteau_2022,Henderson_2014,Stein_2014,Henderson_2015,Kossoski_2021,Marie_2021,Chen_2015,Bytautas_2018}
In particular, coupled cluster restricted to paired double excitations, \cite{Henderson_2014,Stein_2014,Henderson_2015,Kossoski_2021,Marie_2021} which is the same as the antisymmetric product of 1-reference orbital geminals,
\cite{Limacher_2013,Limacher_2014,Tecmer_2014,Boguslawski_2014a,Boguslawski_2015,Boguslawski_2014b,Boguslawski_2014c,Johnson_2017,Fecteau_2020,Johnson_2020,Johnson_2022,Fecteau_2022}
provides very similar energies as DOCI, and at a very favourable polynomial cost.

\begin{figure*}
\includegraphics[width=0.3\linewidth]{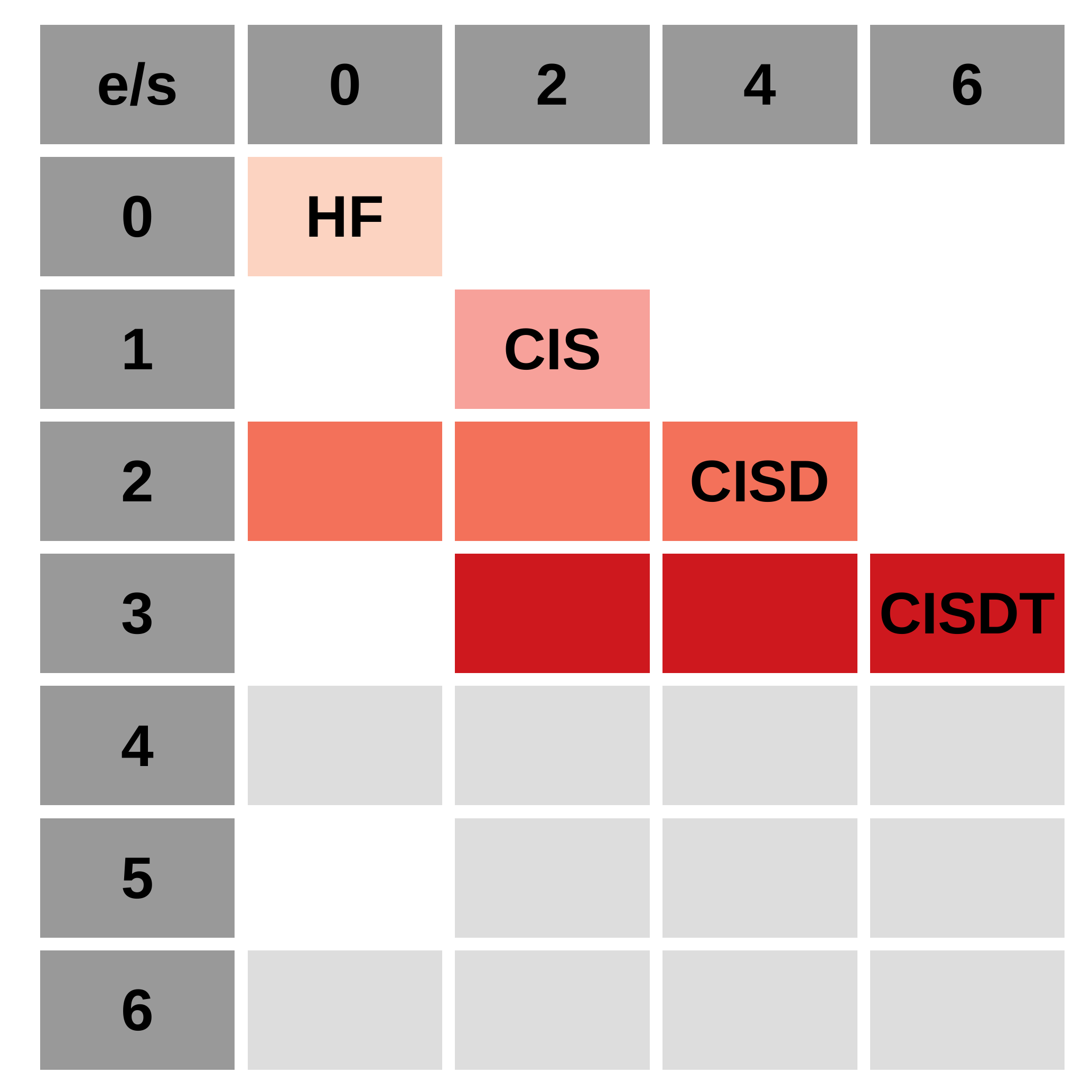}
\hspace{0.02\linewidth}
\includegraphics[width=0.3\linewidth]{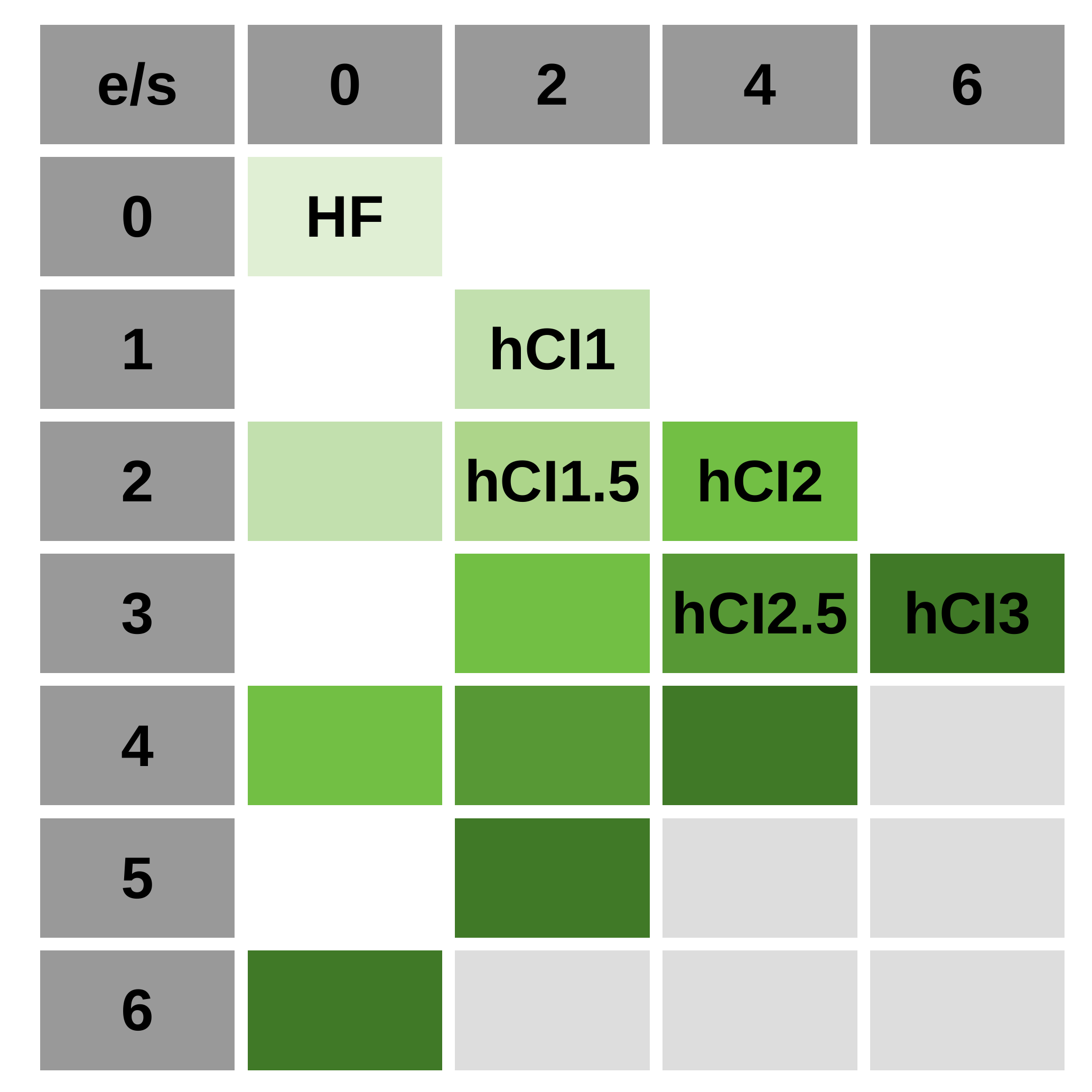}
\hspace{0.02\linewidth}
\includegraphics[width=0.3\linewidth]{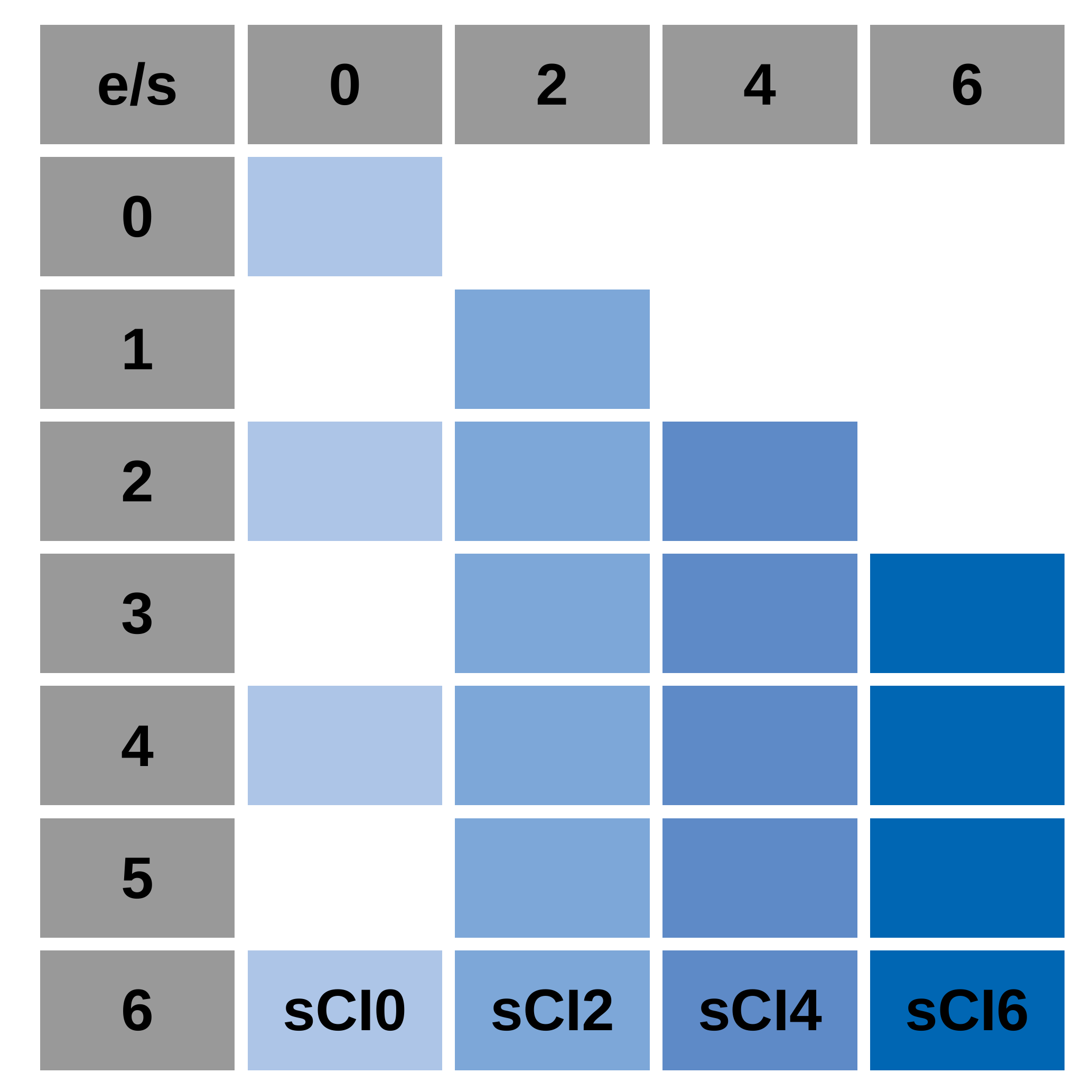}
\caption{Partitioning of the Hilbert space into blocks of specific excitation degree $e$ (with respect to a closed-shell determinant) and seniority number $s$.
This $e$-$s$ map is truncated differently in excitation-based CI (left), seniority-based CI (right), and hierarchy-based CI (center).
The color tones represent the determinants that are included at a given CI level.}
\label{fig:allCI}
\end{figure*}

At this point, we notice the current dichotomy.
When targeting static correlation, seniority-based CI methods tend to have a better performance than excitation-based CI, despite their higher computational cost.
The latter class of methods, in contrast, are well-suited for recovering dynamic correlation, and only at polynomial cost with system size.
Ideally, we aim for a method that captures most of both static and dynamic correlation, with as few determinants as possible.
With this goal in mind, we propose a new partitioning of the Hilbert space, named \textit{hierarchy} CI (hCI).
It combines both the excitation degree $e$ and the seniority number $s$ into one single hierarchy parameter
\begin{equation}
	\label{eq:h}
	h = \frac{e+s/2}{2},
\end{equation}
which assumes half-integer values.
Here we only consider systems with an even number of electrons, meaning that $s$ takes only even values as well.
Figure \ref{fig:allCI} shows how the Hilbert space is progressively populated in excitation-based CI, seniority-based CI, and our hybrid hCI methods.

We have three key justifications for this new CI hierarchy.
The first one is physical.
We know that the lower degrees of excitations and lower seniority sectors, when looked at individually, often carry the most important contribution to the FCI expansion.
By combining $e$ and $s$ as is Eq.~\eqref{eq:h}, we ensure that both directions in the excitation-seniority map (see Fig.~\ref{fig:allCI}) are contemplated.
Rather than filling the map top-bottom (as in excitation-based CI) or left-right (as in seniority-based CI), the hCI methods fills it diagonally.
In this sense, we hope to recover dynamic correlation by moving right in the map (increasing the seniority number while keeping a low excitation degree),
at the same time as static correlation, by moving down (increasing the excitation degree while keeping a low seniority number).
 
The second justification is computational.
In the hCI class of methods, each level of theory accommodates additional determinants from different excitation-seniority sectors (each block of same color tone in Fig.~\ref{fig:allCI}).
The key insight behind hCI is that the number of additional determinants presents the same scaling with respect to $\Nbas$, for all excitation-seniority sectors entering at a given hierarchy $h$.
This justifies the numerator in the definition of $h$ [Eq.~\eqref{eq:h}].
 
Finally, the third justification for our hCI method is empirical and closely related to the computational motivation.
There are many possible ways to populate the Hilbert space starting from a given reference determinant, 
and one can in principle formulate any systematic recipe that includes progressively more determinants.
Besides a physical or computational perspective, the question of what makes for a good recipe can be framed empirically.
Does our hCI class of methods perform better than excitation-based or seniority-based CI,
in the sense of recovering most of the correlation energy with the least computational effort?

Hybrid approaches based on both excitation degree and seniority number have been proposed before. \cite{Alcoba_2014,Raemdonck_2015,Alcoba_2018}
In these works, the authors established separate maximum values for the excitation and the seniority,
and either the union or the intersection between the two sets of determinants have been considered.
For the union case, $\Ndet$ grows exponentially with $\Nbas$,
while in the intersection approach the Hilbert space is filled rectangle-wise in our excitation-seniority map.
In the latter case, the scaling of $\Ndet$ would be dominated by the rightmost bottom block.
Bytautas \textit{et al.}\cite{Bytautas_2015} explored a different hybrid scheme combining determinants having a maximum seniority number and those from a complete active space.
In comparison to previous approaches, our hybrid hCI scheme has two key advantages.
First, it is defined by a single parameter that unifies excitation degree and seniority number [see Eq.~\eqref{eq:h}].
Second and most importantly, each next level includes all classes of determinants whose number share the same scaling with system size, as discussed before, thus preserving the polynomial cost of the method.

Each level of excitation-based CI has a hCI counterpart with the same scaling of $\Ndet$ with respect to $\Nbas$,
justifying the denominator in the definition of $h$ [Eq.~\eqref{eq:h}].
For example, $\Ndet = \order*{\Nbas^4}$ in both hCI2 and CISD, whereas $\Ndet = \order*{\Nbas^6}$ in hCI3 and CISDT, and so on.
From this computational perspective, hCI can be seen as a more natural choice than the traditional excitation-based CI,
because if one can afford for, say, CISDT, then one could probably afford hCI3, due to the same scaling of $\Ndet$.
Of course, in practice an integer-$h$ hCI method has more determinants than its excitation-based counterpart (despite the same scaling of $\Ndet$),
and thus one should first ensure whether including the lower-triangular blocks (going from CISDT to hCI3 in our example)
is a better strategy than adding the next column (going from CISDT to CISDTQ).
Therefore, here we decided to discuss the results in terms of $\Ndet$, rather than the formal scaling of $\Ndet$ as a function of $\Nbas$,
which could make the comparison somewhat biased toward hCI.
It is also interesting to compare the lowest levels of hCI (hCI1) and excitation-based CI (CIS).
Since single excitations do not connect with the reference (at least for HF orbitals), CIS provides the same energy as HF.
In contrast, the paired double excitations of hCI1 do connect with the reference (and the singles contribute indirectly via the doubles).
Therefore, while CIS based on HF orbitals does not improve with respect to the mean-field HF wave function,
the hCI1 counterpart already represents a minimally correlated model, with the same and favorable $\Ndet = \order*{\Nbas^2}$ scaling.
hCI also allows for half-integer values of $h$, with no equivalent in excitation-based CI.
This gives extra flexibility in terms of methodological choice.
For a particular application with excitation-based CI, CISD might be too inaccurate, for example, while the improved accuracy of CISDT might be too expensive.
hCI2.5 could represent an alternative, being more accurate than hCI2 and less expensive than hCI3.

Our main goal here is to assess the performance of hCI against excitation-based and seniority-based CI.
To do so, we have evaluated how fast different observables converge to the FCI limit as a function of $\Ndet$.
In particular, we have calculated the potential energy curves (PECs) for the dissociation of six systems: 
\ce{HF}, \ce{F2}, \ce{N2}, ethylene, \ce{H4}, and \ce{H8},
which display a variable number of bond breaking.
For the latter two molecules, we have considered linearly arranged with equally spaced hydrogen atoms, and computed PECs along the symmetric dissociation coordinate.
For ethylene, we consider the \ce{C=C} double bond breaking, while freezing the remaining internal coordinates.
Its equilibrium geometry was taken from Ref.~\onlinecite{Loos_2018} and is reproduced in the \SupInf.
Due to the (multiple) bond breaking, these are challenging systems for electronic structure methods,
being often considered when assessing novel methodologies.
More precisely, we have evaluated the convergence of four observables: the non-parallelity error (NPE), the distance error, the vibrational frequencies, and the equilibrium geometries.
The NPE is defined as the maximum minus the minimum differences between the PECs obtained at a given CI level and the exact FCI result.
We define the distance error as the maximum plus the minimum differences between a given PEC and the FCI result.
Thus, while the NPE probes the similarity regarding the shape of the PECs, the distance error measures how their overall magnitudes compare.
From the PECs, we have also extracted the vibrational frequencies and equilibrium geometries (details can be found in the \SupInf).


The hCI method was implemented in {\QP} via a straightforward adaptation of the 
\textit{configuration interaction using a perturbative selection made iteratively} (CIPSI) algorithm, \cite{Huron_1973,Giner_2013,Giner_2015,Garniron_2018}
by allowing only for determinants having a given maximum hierarchy $h$ to be selected.
The excitation-based CI, seniority-based CI, and FCI calculations presented here were also performed with the CIPSI algorithm implemented in {\QP}. \cite{Garniron_2019}
It is worth mentioning that the determinant-driven framework of {\QP} allows the inclusion of any arbitrary set of determinants.
In practice, we consider, for a given CI level, the ground state energy to be converged when the second-order perturbation correction computed in the truncated Hilbert space (which approximately measures the error between the selective and complete calculations) lies below \SI{0.01}{\milli\hartree}. \cite{Garniron_2018}
These selected versions of CI require considerably fewer determinants than the formal number of determinants (understood as all those that belong to a given CI level, regardless of their weight or symmetry) of their complete counterparts.
Nevertheless, we decided to present the results as functions of the formal number of determinants (see above), 
which are not related to the particular algorithmic choices of the CIPSI calculations.
The ground-state CI energy is obtained with the Davidson iterative algorithm, \cite{Davidson_1975}
which in the present implementation of {\QP} means that the computation and storage cost us $\order*{\Ndet^{3/2}}$ and $\order*{\Ndet}$, respectively.
This shows that the determinant-driven algorithm is not optimal in general.
However, the selected nature of the CIPSI algorithm drastically reduces the actual number of determinants and therefore calculations are technically feasible.

The CI calculations were performed with both canonical HF orbitals and optimized orbitals.
In the latter case, the energy is obtained variationally in the CI space and in the orbital parameter space, hence defining orbital-optimized CI (oo-CI) methods.
We employed the algorithm described elsewhere \cite{Damour_2021} and also implemented in {\QP} for optimizing the orbitals within a CI wave function.
In order to avoid converging to a saddle point solution, we employed a similar strategy as recently described in Ref.~\onlinecite{Elayan_2022}.
Namely, whenever the eigenvalue of the orbital rotation Hessian is negative and the corresponding gradient component $g_i$ lies below a given threshold $g_0$,
then this gradient component is replaced by $g_0 \abs{g_i}/g_i$.
Here we took $g_0 = $ \SI{1}{\micro\hartree}, and considered the orbitals to be converged when the maximum orbital rotation gradient lies below \SI{0.1}{\milli\hartree}.
While we cannot ensure that the obtained solutions are global minima in the orbital parameter space, we verified that all stationary solutions surveyed here 
correspond to real minima (rather than maxima or saddle points).
All CI calculations were performed with the cc-pVDZ basis set and within the frozen core approximation.
For the \ce{HF} molecule we have also tested basis set effects, by considering the larger cc-pVTZ and cc-pVQZ basis sets.

It is worth mentioning that obtaining smooth PECs for the orbital optimized calculations proved to be far from trivial.
First, the orbital optimization was started from the HF orbitals of each geometry.
This usually led to discontinuous PECs, meaning that distinct solutions were found by our algorithm.
Then, at some geometries that seem to present the lowest lying solution,
the optimized orbitals were employed as the guess orbitals for the neighboring geometries, and so on, until a new PEC is obtained.
This protocol was repeated until the PEC built from the lowest lying oo-CI solution becomes continuous.
We recall that saddle point solutions were purposely avoided in our orbital optimization algorithm. If that was not the case, then even more stationary solutions would have been found.

\begin{figure}[h!]
\includegraphics[width=\linewidth]{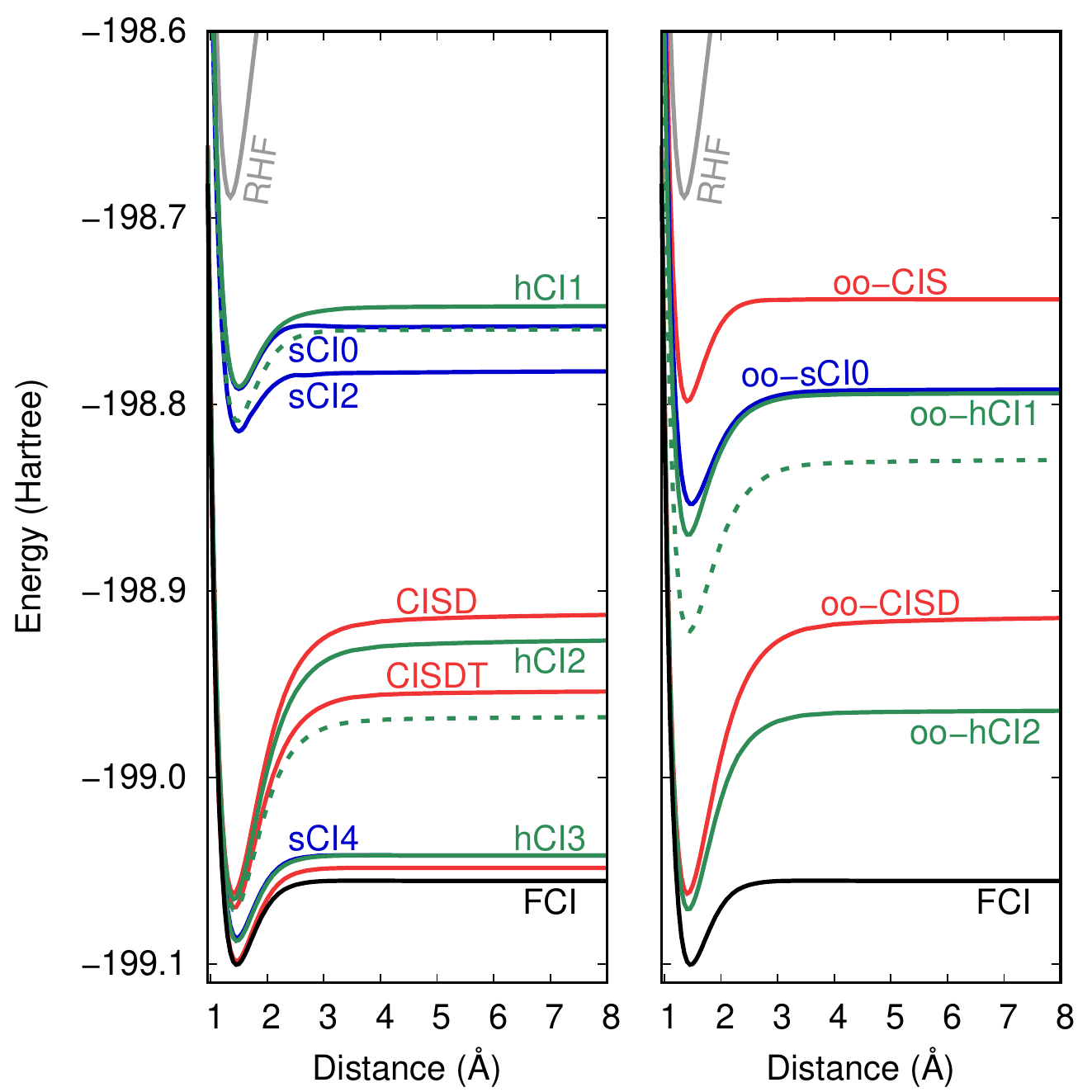}
        \caption{Potential energy curves for \ce{F2},
        according to HF, FCI, and the three classes of CI methods: seniority-based CI (blue), excitation-based CI (red), and hierarchy-based CI (green),
        (dashed lines for half-integer $h$),
        with HF orbitals (left) and orbitals optimized at a given CI level (right),
        and with the cc-pVDZ basis set.}
        \label{fig:F2_pes}
\end{figure}

While the full set of PECs and the corresponding energy differences with respect to FCI are shown in the \SupInf,
in Fig.~\ref{fig:F2_pes} we present the PECs for \ce{F2}, which display many of the features also observed for the other systems.
It already gives a sense of the performance of three classes of CI methods, 
clearly showing the overall superiority of hCI over excitation-based CI.
It further illustrates several important features which will be referenced to in the upcoming discussion.

\begin{figure}[h!]
\includegraphics[width=\linewidth]{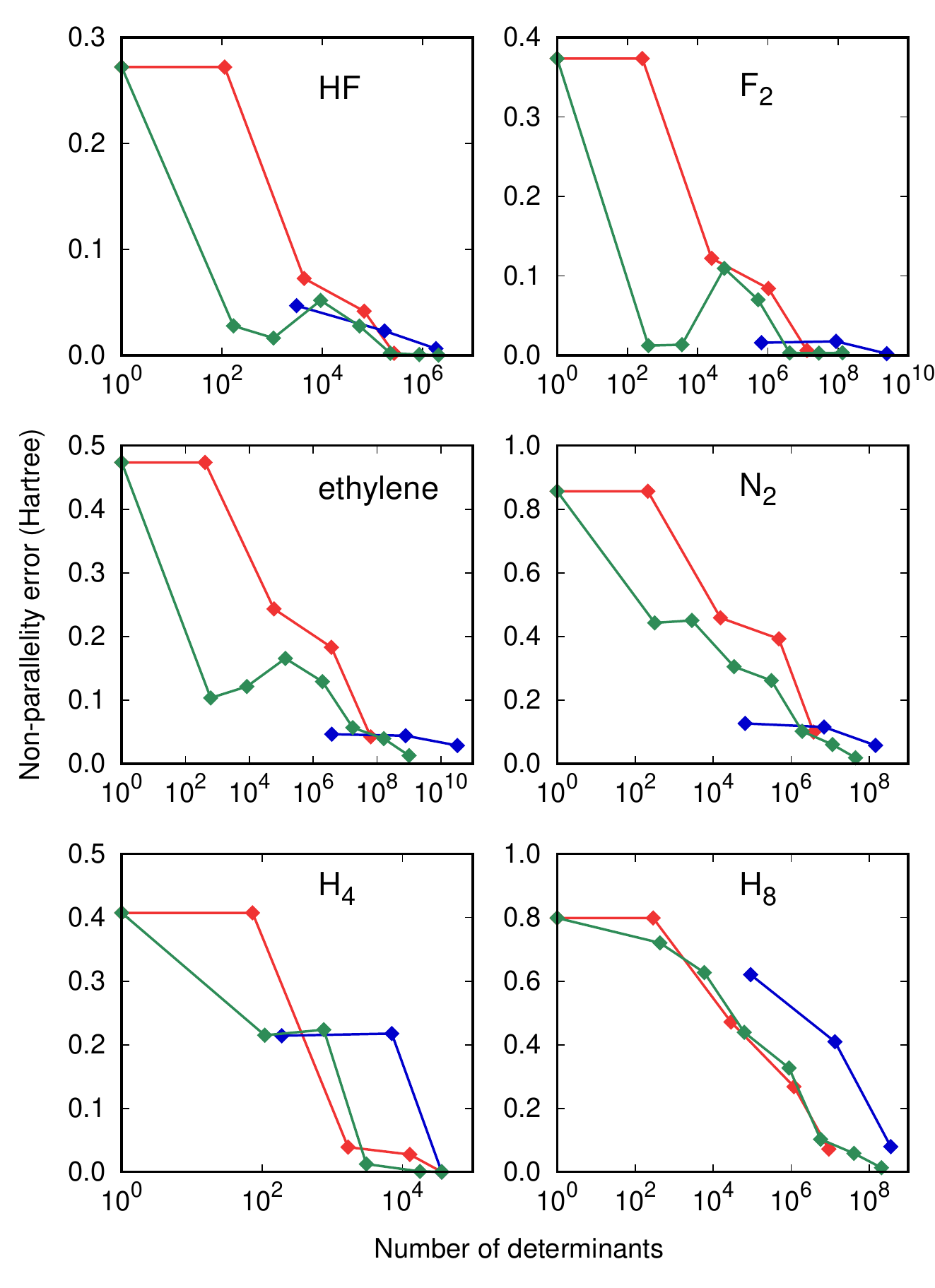}
	\caption{Non-parallelity errors as function of the number of determinants, for the three classes of CI methods: seniority-based CI (blue), excitation-based CI (red), and our proposed hybrid hCI (green).
	}
        \label{fig:plot_stat}
\end{figure}

We first discuss the results for HF orbitals.
In Fig.~\ref{fig:plot_stat}, we present the NPEs for the six systems studied, and for the three classes of CI methods,
as functions of $\Ndet$.
The main result contained in Fig.~\ref{fig:plot_stat} concerns the overall faster convergence of hCI when compared to excitation-based and seniority-based CI.
This is observed for single bond breaking (\ce{HF} and \ce{F2}) as well as the more challenging double (ethylene), triple (\ce{N2}), and quadruple (\ce{H4}) bond breaking.
For \ce{H8}, hCI and excitation-based CI perform similarly.
The convergence with respect to $\Ndet$ is slower in the latter, more challenging cases, irrespective of the class of CI methods, as expected. \cite{Motta_2017,Motta_2020}
But more importantly, the superiority of hCI appears to be highlighted in the one-site multiple bond break systems (compare ethylene and \ce{N2} with \ce{HF} and \ce{F2} in Fig.~\ref{fig:plot_stat}).

For all systems (specially ethylene and \ce{N2}), hCI2 is better than CISD, two methods where $\Ndet$ scales as $\Nbas^4$.
hCI2.5 is better than CISDT (except for \ce{H8}), despite its lower computational cost, whereas hCI3 is much better than CISDT, and comparable in accuracy with CISDTQ (again for all systems).
Inspection of the PECs (see Fig.~\ref{fig:F2_pes} for the case of \ce{F2} or the {\SupInf} for the other systems) reveals that the lower NPEs observed for hCI stem mostly from the contribution of the dissociation region.
This result demonstrates the importance of higher-order excitations with low seniority number in this strong correlation regime, 
which are accounted for in hCI but not in excitation-based CI (for a given scaling of $\Ndet$).
These determinants are responsible for alleviating the size-consistency problem when going from excitation-based CI to hCI.
 
Meanwhile, the first level of seniority-based CI (sCI0, which is the same as DOCI)
tends to offer a rather low NPE when compared to the other CI methods with a similar $\Ndet$ (hCI2.5 and CISDT).
However, convergence is clearly slower for the next levels (sCI2 and sCI4), whereas excitation-based CI and specially hCI converge faster.
Furthermore, seniority-based CI becomes less attractive for larger basis set in view of its exponential scaling.
This can be seen in Figs.~S2 and S3 of the \SupInf, which shows that augmenting the basis set leads to a much steeper increase of $\Ndet$ for seniority-based CI.

It is worth mentioning the surprisingly good performance of hCI1 and hCI1.5.
For \ce{HF}, \ce{F2}, and ethylene, they yield lower NPEs than the much more expensive CISDT method, and only slightly higher in the case of \ce{N2}.
For the same systems, we also see the NPEs increase from hCI1.5 to hCI2, and decreasing to lower values only at the hCI3 level. 
(Even than, it is important to remember that the hCI2 results remain overall superior to their excitation-based counterparts.)
Both findings are not observed for \ce{H4} and \ce{H8}.
It seems that both the relative worsening of hCI2 and the success of hCI1 and hCI1.5
become less apparent as progressively more bonds are being broken (compare, for instance, \ce{F2}, \ce{N2}, and \ce{H8} in Fig.~\ref{fig:plot_stat}).
This reflects the fact that higher-order excitations are needed to properly describe multiple bond breaking,
and also hints at some cancelation of errors in low-order hCI methods for single bond breaking.

In Fig.~S5 of the \SupInf, we present the distance error, which is also found to decrease faster with hCI.
Most of the observations discussed for the NPE also hold for the distance error, with two main differences.
The convergence is always monotonic for the latter observable (which is expected from its definition),
and the performance of seniority-based CI is much poorer (due to the slow recovery of dynamic correlation).

In Figs.~S6 and S7 of the \SupInf, we present the convergence of the equilibrium geometries and vibrational frequencies, respectively,
as functions of $\Ndet$, for the three classes of CI methods.
For the equilibrium geometries, hCI performs slightly better overall than excitation-based CI.
A more significant advantage of hCI can be seen for the vibrational frequencies.
For both observables, hCI and excitation-based CI largely outperform seniority-based CI.
Similarly to what we have observed for the NPEs, the convergence of hCI is also found to be non-monotonic in some cases.
This oscillatory behavior is particularly evident for \ce{F2} 
(notice in Fig.~\ref{fig:F2_pes} how the potential well goes from shallow in hCI1, to deep in hCI2, and then shallow again in hCI3).
It is also noticeable for \ce{HF}, becoming less apparent for ethylene, virtually absent for \ce{N2},
and showing up again for \ce{H4} and \ce{H8}.
Interestingly, equilibrium geometries and vibrational frequencies of \ce{HF} and \ce{F2} (single bond breaking), 
are rather accurate when evaluated at the hCI1.5 level, bearing in mind its relatively modest computational cost.

For the \ce{HF} molecule we have also evaluated how the convergence is affected by increasing the size of the basis set, going from cc-pVDZ to cc-pVTZ and cc-pVQZ (see Figs.~S2 and S3 in the \SupInf).
While a larger $\Ndet$ is required to achieve the same level of convergence, as expected,
the convergence profiles remain very similar for all basis sets.
Vibrational frequency and equilibrium geometry present less oscillations for hCI.
We thus believe that the main findings discussed here for the other systems would be equally basis set independent.

Up to this point, all results and discussions have been based on CI calculations with HF orbitals.
We recall that seniority-based CI (in contrast to excitation-based CI) is not invariant with respect to orbital rotations within the occupied and virtual subspaces, \cite{Bytautas_2011}
and for this reason it is customary to optimize the corresponding wave function by performing such rotations.
Similarly, hCI wave functions are not invariant under orbital rotations within each subspace.
Thus, we decided to further assess the role of orbital optimization (occupied-virtual rotations included) for each class of CI methods.
Due to the significantly higher computational cost and numerical difficulties associated with orbital optimization at higher CI levels,
such calculations were typically limited up to oo-CISD (for excitation-based), oo-DOCI (for seniority-based), and oo-hCI2 (for hCI).
The PECs and convergence of properties as function of $\Ndet$ are shown in the \SupInf.

Of course, at a given CI level, orbital optimization will lead to lower energies than with HF orbitals. 
However, even though the energy is lowered (thus improved) at each geometry, such improvement may vary largely along the PEC, which may or may not decrease the NPE.
More often than not, the NPEs do decrease upon orbital optimization, though not always.
For example, compared with their non-optimized counterparts, oo-hCI1 and oo-hCI1.5 provide somewhat larger NPEs for \ce{HF} and \ce{F2},
similar NPEs for ethylene, and smaller NPEs for \ce{N2}, \ce{H4}, and \ce{H8}.
Following the same trend, oo-CISD presents smaller NPEs than HF-CISD for the multiple bond breaking systems, but very similar ones for the single bond breaking cases.
oo-CIS has significantly smaller NPEs than HF-CIS, being comparable to oo-hCI1 for all systems except for \ce{H4} and \ce{H8}, where the latter method performs better.
(We will come back to oo-CIS later.)
Based on the present oo-CI results, hCI still has the upper hand when compared with excitation-based CI, though by a smaller margin.

Orbital optimization usually reduces the NPE for seniority-based CI (in this case we only considered oo-DOCI) as well.
The gain is specially noticeable for \ce{H4} and \ce{H8} (where the orbitals become symmetry-broken \cite{Henderson_2014}),
and much less so for \ce{HF}, ethylene, and \ce{N2} (where the orbitals remain symmetry-preserved).
This is in line with what has been observed before for \ce{N2}. \cite{Bytautas_2011}
For \ce{F2}, we found that orbital optimization actually increases the NPE (though by a small amount),
due to the larger energy lowering in the Franck-Condon region than at dissociation (see Fig.~\ref{fig:F2_pes}).
These results suggest that, when bond breaking involves one site, orbital optimization at the DOCI level does not have such an important role,
at least in the sense of decreasing the NPE.

Optimizing the orbitals at the CI level also tends to benefit the convergence of vibrational frequencies and equilibrium geometries.
The impact is often somewhat larger for hCI than for excitation-based CI, by a small margin.
Also, the large oscillations observed in the hCI convergence with HF orbitals (for \ce{HF} and \ce{F2}) are significantly suppressed upon orbital optimization.

We come back to the surprisingly good performance of oo-CIS, which is interesting due to its low computational cost.
The PECs are compared with those of HF and FCI in Fig.~S12 of the \SupInf.
At this level, the orbital rotations provide an optimized reference (different from the HF determinant), from which only single excitations are performed.
Since the reference is not the HF determinant, Brillouin's theorem no longer holds, and single excitations actually connect with the reference.
Thus, with only single excitations (and a reference that is optimized in the presence of these excitations), one obtains a minimally correlated model.
Interestingly, oo-CIS recovers a non-negligible fraction (15\%-40\%) of the correlation energy around the equilibrium geometries.
For all systems, significantly more correlation energy (25\%-65\% of the total) is recovered at dissociation.
In fact, the larger account of correlation at dissociation is responsible for the relatively small NPEs encountered at the oo-CIS level.
We also found that the NPE drops more significantly (with respect to the HF one) for the single bond breaking cases (\ce{HF} and \ce{F2}), 
followed by the double (ethylene) and triple (\ce{N2}) bond breaking, then \ce{H4}, and finally \ce{H8}.

The above findings can be understood by looking at the character of the oo-CIS orbitals.
At dissociation, the closed-shell reference is actually ionic, with orbitals assuming localized atomic-like characters.
The reference has a decreasing weight in the CI expansion as the bond is stretched, becoming virtually zero at dissociation.
However, that is the reference one needs to achieve the correct open-shell character of the fragments when the single excitations of oo-CIS are accounted for.
Indeed, the most important single excitations promote the electron from the negative to the positive fragment, resulting in two singly open-shell radicals.
This is enough to obtain the qualitatively correct description of single bond breaking, hence the relatively low NPEs observed for \ce{HF} and \ce{F2}.
In contrast, the oo-CIS method can only explicitly account for one unpaired electron on each fragment, such that multiple bond breaking become insufficiently described.
Nevertheless, double (ethylene) and even triple (\ce{N2}) bond breaking still appear to be reasonably well-described at the oo-CIS level.


In this Letter, we have proposed a new scheme for truncating the Hilbert space in configuration interaction calculations, named hierarchy CI (hCI).
By merging the excitation degree and the seniority number into a single hierarchy parameter $h$, 
the hCI method ensures that all classes of determinants sharing the same scaling of $\Ndet$ with the number of basis functions are included in each level of the hierarchy.
We evaluated the performance of hCI against excitation-based CI and seniority-based CI,
by comparing PECs and derived quantities 
for six systems, ranging from single to multiple bond breaking.

Our key finding is that the overall performance of hCI either surpasses or equals that of excitation-based CI,
in the sense of convergence with respect to $\Ndet$.
The superiority of hCI is more noticeable for the non-parallelity and distance errors, but also observed to a lesser extent for the vibrational frequencies and equilibrium geometries.
The comparison to seniority-based CI is less trivial.
DOCI (the first level of seniority-based CI) often provides even lower NPEs for a similar $\Ndet$, but it falls short in describing the other properties investigated here.
In addition, if higher accuracy is desired, convergence was found to be faster with hCI (and also excitation-based CI) than seniority-based CI, at least for HF orbitals.
Finally, the exponential scaling of seniority-based CI in practice precludes this approach for larger systems and basis sets,
while the favorable polynomial scaling and encouraging performance of hCI is an alternative.
 
We found surprisingly good results for the first level of hCI (hCI1) and the orbital optimized version of CIS (oo-CIS), two methods with very favorable computational scaling.
In particular, oo-CIS correctly describes single bond breaking.
We hope to report on generalizations to excited states in the future.
In contrast, orbital optimization at higher CI levels is not necessarily a recommended strategy, 
given the overall modest improvement in convergence when compared to results with canonical HF orbitals.
One should bear in mind that optimizing the orbitals is always accompanied with well-known challenges (several solutions, convergence issues, etc)
and may imply a significant computational burden (associated with the calculations of the orbital gradient and Hessian, and the many iterations that are often required),
specially for larger CI spaces.
In this sense, stepping up in the CI hierarchy might be a more straightforward and possibly a cheaper alternative than optimizing the orbitals.
One possibility to explore is to first optimize the orbitals at a lower level of CI, and then to employ this set of orbitals at a higher level of CI.

The hCI pathway presented here offers several interesting possibilities to pursue.
One could generalize and adapt hCI for excited states \cite{Veril_2021} and open-shell systems, \cite{Loos_2020}
develop coupled-cluster methods based on an analogous excitation-seniority truncation of the excitation operator, \cite{Aroeira_2021,Magoulas_2021,Lee_2021}
and explore the accuracy of hCI trial wave functions for quantum Monte Carlo simulations. \cite{Dash_2019,Dash_2021,Cuzzocrea_2022}

\begin{acknowledgements}
This work was performed using HPC resources from CALMIP (Toulouse) under allocation 2021-18005.
This project has received funding from the European Research Council (ERC) under the European Union's Horizon 2020 research and innovation programme (Grant agreement No.~863481).
\end{acknowledgements}

\section*{Supporting information available}
\label{sec:SI}
Potential energy curves, energy differences with respect to FCI results, non-parallelity errors, distance errors, vibrational frequencies, and equilibrium geometries,
according to the three classes of CI methods (excitation-based CI, seniority-based CI, and hierarchy-based CI),
with Hartree-Fock orbitals and with optimized orbitals,
computed for \ce{HF}, \ce{F2}, ethylene, \ce{N2}, \ce{H4}, and \ce{H8}, with the cc-pVDZ basis set,
and also for \ce{HF}, with cc-pVDZ, cc-pVTZ, cc-pVQZ basis sets.
Equilibrium geometry of ethylene,
and details about the fitting of potential energy curves.


\bibliography{seniority}

\end{document}